\documentclass[aps,twocolumn,prl,showpacs]{revtex4}

\input{tcilatex}

\begin{document}

\title{Formally exact quantization condition for nonrelativistic quantum
systems}
\author{Yong-Cheng \surname{Ou}}
\email{ouyongcheng@163.com}

\affiliation{Department of Physics, Shanghai Jiao Tong University,
Shanghai 200240, People's Republic of China}

\author{Zhuang-Qi Cao, and Qi-Shun Shen}

\affiliation{Department of Physics, Shanghai Jiao Tong University,
Shanghai 200240, People's Republic of China}

\begin{abstract}
Based on the standard transfer matrix, a formally exact quantization
condition for arbitrary potentials, which outflanks and unifies the
historical approaches, is derived. It can be used to find the exact
bound-state energy eigenvalues of the quantum system without solving an
equation of motion for the system wave functions.
\end{abstract}

\maketitle

In general an exact solution for many quantum mechanical problem is
unavoidable and one is forced to resort to some types of approximation
technique. The Wentzel-Kramers-Brillouin(WKB) approximation, based on the
wave function expansion in powers of $\hbar $, is studied and applied
frequently in many fields. Apparently, it only remains valid for the
short-waves and then plays a role of a bridge between classical and quantum
mechanics. Neglecting of the higher order of $\hbar $ directly leads to
failing to describe the wave function at the classical turning points,
together with inaccurate energy eigenvalues in the lower bound states. Langer%
\cite{langer} introduced an approximation based on Bessel function, coming
across the same problems as the WKB approximation. Rather than in terms of
the exponential function, Miller\cite{miller} proposed a method by proper
choice of the arbitrary equation. Friedrich\cite{firedrich} believed the
phase shift at the tuning point is energy-dependent with nonintegral Maslov
indices in the limit of the long-waves. Recently, the structure of
supersymmetric quantum mechanics motivates a modified semiclassical
quantization condition for one-dimensional Hamiltonians. But the
supersymmetric WKB(SWKB) method only reproduces the exact energy eigenvalues
for translationally shape-invariant potentials whose ground-state wave
function is analytically known\cite{comtet} \cite{barclay}. When the two or
three-dimensional potentials are dealt with, Langer-like corrections \cite%
{langer} is needed, which is exact only for solvable spherically symmetric
potentials. All above methods are unable to give an exact quantization
condition for arbitrary nonrelativistic quantum systems.

In this paper, from a different starting point, we first take the standard
transfer matrix as the basis of analysis, and deduce the exact quantization
condition in which the scattering-led phase shift recovers the neglecting of
series of higher order $\hbar $ in the WKB approximation. However, the
scattering-led phase shift term is of discrete sum formalism, furthermore,
it is a bit purer mathematical and lacks of physical insights. But the
defined equivalent attenuation vector is recognized to be nothing more than $%
-\psi ^{^{\prime }}\left( x\right) /\psi \left( x\right) ,$ so that an
elegant integral formalism of the scattering-led phase shift can be instead
of the discrete sum formalism. A Riccati equation is obtained, which is
associated the wave function with the scattering-led phase shift, and we
show the quantization condition outflanks and unifies the historical
approaches, leading directly to the exact eigenvalues and the exact
eigenfunction so that we can define an exact real quantum action-angle
variable $J\left( E\right) $ determining the energy eigenvalues.

Without loss of generality, we assume a potential with two classical turning
points $x_{L}$ and $x_{R}$ which are determined from $V\left( x_{L}\right)
=V\left( x_{R}\right) =E$. Far from the two points, we truncate the
potential at $x_{C}$ and $x_{D}$. In order to employ the transfer matrix, $%
\left[ x_{C},x_{D}\right] $ is divided into $l+f+g$ equal section layers
with width $h$, letting $\left[ x_{C},x_{L}\right] $ have $l$ section
layers, $\left[ x_{L},x_{R}\right] $ have $f$ section layers and $\left[
x_{R},x_{D}\right] $ have $g$ section layers. In each section layer, the
potential can be regarded as constant. Hence the transfer matrix of $j$ th
layer can be written as
\begin{equation}
M_{j}=\left[
\begin{array}{cc}
\cos \left( \kappa _{j}h\right)  & -\frac{1}{\kappa _{j}}\sin (\kappa _{j}h)
\\
\kappa _{j}\sin (\kappa _{j}h) & \cos (\kappa _{j}h)%
\end{array}%
\right] ,
\end{equation}%
where $k_{j}$ is defined as $\hbar \kappa _{j}=\sqrt{2m\left[ E-V_{j}\right]
}$ and the subscript $j$ is an integer ranging from 1 to $l+f+g$. $V_{j}$
stands for the potential energy at the $j$ th section layer, $m$ is the
particle mass and $E$ is the energy eigenvalue. Using the boundary condition
that the wave function $\psi \left( x\right) $ and its first derivative $%
\psi ^{\prime }\left( x\right) $ are continuous at the boundary between two
neighboring section layers, the matrix equation is given
\begin{equation}
\left[
\begin{array}{c}
\psi \left( x_{C}\right)  \\
\psi ^{\prime }\left( x_{C}\right)
\end{array}%
\right] =\prod_{j=1}^{l+f+g}M_{j}\left[
\begin{array}{c}
\psi \left( x_{D}\right)  \\
\psi ^{\prime }\left( x_{D}\right)
\end{array}%
\right] .
\end{equation}%
The right- and left-hand side in Eq.(2) are simultaneously multiplied by the
matrix $\left[ -\psi ^{^{\prime }}\left( x_{C}\right) ,\psi \left(
x_{C}\right) \right] $ and divided by $\psi \left( x_{C}\right) \psi \left(
x_{D}\right) $, then Eq.(2) is changed to the form
\begin{equation}
\left[
\begin{array}{cc}
-\frac{\psi ^{\prime }\left( x_{C}\right) }{\psi \left( x_{C}\right) } & 1%
\end{array}%
\right] \prod_{j=1}^{l+f+g}M_{j}\left[
\begin{array}{c}
1 \\
\frac{\psi ^{\prime }\left( x_{D}\right) }{\psi \left( x_{D}\right) }%
\end{array}%
\right] =0.
\end{equation}%
The following analysis is based on the section layer width $h\rightarrow 0$,
the mass $m=1$ and Planck's constant $\hbar =1$. We define $P_{C}=-\psi
^{\prime }\left( x_{C}\right) /\psi \left( x_{C}\right) $ and $P_{D}=\psi
^{\prime }\left( x_{D}\right) /\psi \left( x_{D}\right) $. Making use of the
transfer characteristic of matrix in Eq.(3), the values of $-\psi ^{\prime
}\left( x\right) /\psi \left( x\right) $ are calculated at any position $x$.
After algebraic manipulations\cite{cao}, Eq.(3) is equivalent to
\begin{eqnarray}
&&\sum_{w=l+1}^{l+f}\kappa _{w}h+\delta \left( n\right)   \nonumber \\
&=&n\pi +\arctan \left( \frac{P_{l}}{\kappa _{l+1}}\right) +\arctan \left(
\frac{P_{l+f+1}}{\kappa _{l+f}}\right) ,
\end{eqnarray}%
where $\delta \left( n\right) $ is determined from
\begin{equation}
\delta \left( n\right) =\sum_{i=l+1}^{l+f-1}\left[ \arctan \left( \frac{%
P_{i+1}}{\kappa _{i+1}}\right) -\arctan \left( \frac{P_{i+1}}{\kappa _{i}}%
\right) \right] ,  \label{9}
\end{equation}%
with
\begin{equation}
P_{j}=\kappa _{j}\left[ \arctan \left( \frac{P_{j+1}}{\kappa _{j}}\right)
-\kappa _{j}h\right] ,  \label{11}
\end{equation}%
and $P_{j}$ are subject to the boundary condition $%
P_{0}=P_{C},P_{l+f+g+1}=P_{D}$. From Eq.(5), we know that if the potential
is nonconstant, $\delta \left( n\right) $ is nonzero, so that $\delta \left(
n\right) $ can be interpreted as the scattering-led phase shift which is in
general energy-dependent or quantum number $n$-dependent. It is clear that $%
\kappa _{l+1}=\sqrt{2\left[ E-V_{l+1}\right] }\rightarrow 0$ and $\kappa
_{l+f}=\sqrt{2\left[ E-V_{l+f+1}\right] }\rightarrow 0$, while $P_{l}$ and $%
P_{l+f+1}$ is positive and finite for bound states, so $\arctan {\left(
P_{l}/\kappa _{l+1}\right) }=\arctan {\left( P_{l+f+1}/P_{l+f}\right) }=\pi
/2$ that is the energy-independent phase shift at the turning points. If a
wave is reflected at $x_{L}$ where $V\left( x_{L}\right) \rightarrow \infty $%
, meanwhile, $P_{l}\rightarrow +\infty $ and $\kappa _{l+1}$ is a finite
positive number, so the half phase shift at $x_{L}$ $\arctan \left(
P_{l}/\kappa _{l+1}\right) $ is also exactly equal to the same $\pi /2$ as
that at the turning point. The first term in Eq.(4) can be integrated as $%
\sum_{w=l+1}^{l+f}\kappa _{w}h=\int_{x_{L}}^{x_{R}}\kappa dx$. Given that
Eq.(6) is equivalent to $\kappa _{j}h=\arctan {\left( P_{j+1}/\kappa
_{j}\right) }-\arctan {\left( P_{j}/\kappa _{j}\right) }$, and the
derivative of $\arctan {\left( P/\kappa \right) }$, namely, $d\left[ \arctan
{\left( P/\kappa \right) }\right] /dx=\left( \kappa dP-Pd\kappa \right)
/\left( \kappa ^{2}+P^{2}\right) $, we have
\begin{equation}
\arctan \left( \frac{P_{j+1}}{\kappa _{j}}\right) -\arctan \left( \frac{P_{j}%
}{\kappa _{j}}\right) =\frac{\Delta P\kappa _{j}-P_{j}\Delta \kappa }{\kappa
_{j}^{2}+P_{j}^{2}}.
\end{equation}%
Due to $\Delta \kappa =\kappa _{j}-\kappa _{j}=0$ and $\Delta P=P_{j+1}-P_{j}
$, Eq.(7) reduces to a first order differential equation
\begin{equation}
\frac{dP\left( x\right) }{dx}=\kappa ^{2}\left( x\right) +P^{2}\left(
x\right) ,
\end{equation}%
where
\begin{equation}
P\left( x\right) =-\frac{\psi ^{^{\prime }}\left( x\right) }{\psi \left(
x\right) },
\end{equation}%
which is a well-known Riccati equation. Substitution Eq.(9) into Eq.(8)
results in the time-independent Schr\"{o}dinger equation $\psi ^{^{\prime
\prime }}\left( x\right) +\kappa ^{2}\psi \left( x\right) =0$. From another
angle, the deducing of Eq.(8) is proven to be true. Likewise, Eq.(5) can be
integrated as
\begin{equation}
\delta \left( n\right) =\int_{x_{L}}^{x_{R}}\left( -\frac{d\kappa }{dx}\frac{%
P}{P^{2}+\kappa ^{2}}\right) dx=\int_{x_{L}}^{x_{R}}\left( -\kappa
^{^{\prime }}\frac{P}{P^{^{\prime }}}\right) dx.
\end{equation}%
where $\kappa ^{^{\prime }}=d\kappa /dx$ and $P^{^{\prime }}=dP/dx$. From
Eq.(9), we obtain the wave function
\begin{equation}
\psi \left( x\right) =N_{0}\exp {\left[ -\int^{x}P\left( x^{^{\prime
}}\right) dx^{^{\prime }}\right] },
\end{equation}%
which is acquired to be square integrable. In order to determine $P\left(
x\right) $, one must solve Eq.(8). The initial condition of Eq.(8) is
determined from $P_{D}$, while the function $\psi \left( x\right) =N_{1}\exp
{\left[ -\alpha \left( x-x_{D}\right) \right] }$ is matched to the wave
function decaying exponentially in the range $\left[ x_{D},\infty \right] $,
with $\alpha =\sqrt{2m\left[ V\left( x_{D}\right) -E\right] }$ which is the
equivalent attenuated vector, so $P_{D}=-\psi ^{^{\prime }}\left(
x_{D}\right) /\psi \left( x_{D}\right) =\alpha $.

Now from Eq.(4) and Eq.(10), i.e., $\int_{x_{L}}^{x_{R}} \kappa dx+\delta
\left(n\right)=\left(n+1\right)\pi,$ we have derived the exact quantization
condition
\begin{equation}
\oint_{C} \left(\kappa-\kappa^{^{\prime}} \frac{P}{P^{^{\prime}}}%
\right)dx=2\left(n+1\right)\pi,n=0,1,2...,
\end{equation}
where the integral is counterclockwise around a closed contour $C$ which
encloses the two turning points. The quantization condition complies with
the requiring the total phase during one period of oscillation to be an
integral multiple of $2\pi$ for a particle oscillating in the classically
allowed region between the two turning points $\left[x_{L},x_{R}\right]$.
Comparing with scattering theory in which the sub-waves phase is shifted
when the particle is scattered by a potential, we can gain the better
understanding why there is a term $\delta\left(n\right)$ in Eq.(12), which
is ignored by other approximations such as the WKB approximation and the
Bohr-Sommerfeld quantization condition.

In contrast with the classical Hamilton-Jacobi theory, the quantum
action-angle variable eigenvalues $J$ should be defined as $J=J\left(
E\right) =\left( 1/2\pi \right) \oint_{C}Kdx$ with $K=\kappa -\kappa
^{^{\prime }}P/P^{^{\prime }}$. Since $-\kappa ^{^{\prime }}P/P^{^{\prime }}$
has the same momentum dimension as $\kappa $, $K$ is referred to as the
complete quantum momentum function(QMF). The point is emphasized that the $%
J\left( E\right) $ is defined at arbitrary energies, not only the
eigenvalues. The wave function outside the classically allowed region,
corresponding to whether the energy eigenvalues or other arbitrary energies,
cannot be exponential growth, on the contrary, is bound to decay
exponentially because the potential $V(x)$ is larger than the particle $E$
outside the classically allowed region, the particle cannot fully enter the
classically forbidden region, implying the wave function should decay
exponentially. As a result, there is no the points where log derivative $P$
diverges. Considering the exponentially decaying wave function outside the
classically allowed region and with the help of the Riccati Eq.(8), $J(E)$
in Eq.(12) can be calculated for arbitrary $E$, clearly the behavior of $%
J(E) $ is monotonic in all $E$. The new definition of $J$ is a conceptual
breakthrough not only because no one reached it from the founding of quantum
mechanics to now but also Eq.(12) possesses the basis going back to the
Bohr-Sommerfeld quantization condition in the old quantum theory. Thereby, a
one-to-one correspondence between the invariant torus and quantum
eigenvalues is unambiguously established.

It should be noted that our defined $J\left( E\right) $ is a real function
which describes the system motion, and, in addition, enables one to complete
Hamilton's original program of associating a wave with a particle motion. In
essence, the scattering of a particle in the bound state alters the momentum
of the particle, as a result, contributes significantly to the exactness of
Eq.(12). At the same time, Eq.(12) indicates the energy eigenvalue is
involved both with the particle behavior between the two turning points and
with the particle behavior outside the two turning points, which is the
different aspect between the classical and quantum mechanics, while the
condition $J\left( E\right) =n+1$ yields the exact energy eigenvalues, the
number 1 comes from the contribution due to the phase shift at the turning
points. In view of the Einstein-Brillouin-Keller (EBK) quantization condition%
\cite{sch} \cite{33}, since it only includes the particle behavior between
the two turning points, it is impossible to generally give birth to the
exact results.

It is well worth mentioning the Bohm's hidden variable theory\cite{bohm}. A
pair of equations is obtained, in which one of terms is called quantum
potential, while we define a QMF $K$ in Eq.(12). Apparently whether the
so-called quantum potential or the QMF is related to the wavefunction.
According to Bohm's theory Newton's second law can be written as $md^{2}%
\mathbf{r}/dt^{2}=-\nabla \left( V+Q\right) |_{r=r\left( t\right) }$\cite{de}%
, where $V$ is classical potential and $Q$ is the quantum potential. But in
our theory the classical $\kappa $ and quantum momentum $-\kappa ^{^{\prime
}}P/P^{^{\prime }}$, namely, the QMF $K$, contribute to the quantum dynamics
of the system described by $md\mathbf{r}/dt=\kappa -\kappa ^{^{\prime
}}P/P^{^{\prime }}|_{r=r\left( t\right) }$. This detailed investigation will
appear elsewhere.

The quantization condition Eq.(12) together with Eq.(8)-Eq.(11) is exact
both for the solvable potentials and for the unsolvable potentials, in order
to illustrate them explicitly, let us first examine the detailed values of
the scattering-led phase shift and detailed expressions of the phase
integral for several familiar solvable potentials. Why these potentials are
chosen is that their analytical wave functions are known so that one can
calculate $\delta\left(n\right)$, consequently, one can confirm the
exactness of Eq.(8)-Eq.(12).

(1). One-dimensional infinite square well $V\left(x\right)=0$ at $x\subseteq %
\left[0,L\right] $ but $\infty $ at other $x$. Due to $d\kappa/dx=0$, one
has $\delta\left(n\right)=0$ and the quantization condition reads
\begin{equation}
\int_{0}^{L}\sqrt{2E}dx=\left(n+1\right)\pi, n=0,1,2...,
\end{equation}
which yields exact energy eigenvalues $E_{n}=\left(n+1\right)^{2}E_{0}$ with
$E_{0}=\pi^{2}\hbar^{2}/2L^{2}$. However, the WKB approximation $\int_{0}^{L}%
\sqrt{2E}dx=\left(n+1/2\right)\pi\hbar$ can not do so.

(2). One-dimensional harmonic oscillator $V\left( x\right) =x^{2}/2$, whose
eigenfunction can be expressed by elementary functions, i.e., $\psi _{0}=%
\sqrt{\alpha }/\pi ^{1/4}\exp {\left[ -\alpha ^{2}x^{2}/2\right] }$, $\psi
_{1}=\sqrt{2\alpha }/\pi ^{1/4}\alpha x\exp {\left[ -\alpha ^{2}x^{2}/2%
\right] }$..., substituting $\psi _{n}$ into Eq.(8) and Eq.(10), we find $%
\delta \left( n\right) =\pi /2$, to a certain degree, which shows Eq.(10) is
exact because Eq.(12) reduces to the familiar WKB quantization condition
that remains exact for the one-dimensional harmonic oscillator. For the
three-dimensional harmonic oscillator $V_{eff}\left( r\right)
=r^{2}/2+l\left( l+1\right) /2r^{2}$, substituting the known wave functions
into Eq.(8) and Eq.(10), we have $\delta \left( n,l\right) =\left[ 2\sqrt{%
l\left( l+1\right) }-\left( 2l-1\right) \right] /4$, and the quantization
condition reads
\begin{equation}
\int_{r_{L}}^{r_{R}}\kappa dr+\frac{2\sqrt{l\left( l+1\right) }-\left(
2l-1\right) }{4}\pi =\left( n+1\right) \pi,
\end{equation}%
which yields the exact energy eigenvalues $E=2n+l+3/2$. $\kappa $ is defined
by $\kappa =\sqrt{2m\left[ E-V_{eff}\left( r\right) \right] }$ in Eq.(14).
For this case, the total Maslov index\cite{firedrich} $\mu $ happens to
accord with the term $2l-2\sqrt{l\left( l+1\right) }+3$.
\begin{table}[tbp]
\begin{tabular}{ccccc}
\hline\hline
$n$ & $E^{exact}$ & $E^{present}$ & $E^{Maslov}$ &  \\ \hline
0 & -0.97815416 & -0.97815416 & -0.97834291 &  \\
1 & -0.93556613 & -0.93556613 & -0.93566866 &  \\
2 & -0.87203511 & -0.87203511 & -0.87210568 &  \\
3 & -0.78795362 & -0.78795362 & -0.78800723 &  \\
4 & -0.68386490 & -0.68386490 & -0.68390852 &  \\
5 & -0.56051533 & -0.56051533 & -0.56055291 &  \\
6 & -0.41901295 & -0.41901295 & -0.41904703 &  \\
7 & -0.26131274 & -0.26131274 & -0.09251698 &  \\
8 & -0.09248716 & -0.09248716 & -0.09251698 &  \\ \hline\hline
\end{tabular}%
\caption{Comparisons of exact, present, and the Maslov index energy
eigenvalues for the radial Woods-Saxon potential $V\left(r\right)=-\frac{1}{%
1+\exp{\left[2\left(r-30\right)\right]}} +\frac{l\left(l+1\right)}{2r^2}$
with $l=1$, setting $m=\hbar =1$.}
\end{table}
(3). One-dimensional Coulomb potential $V\left( x\right) =-1/|x|$. Though
there is a singularity at $x=0$ \cite{jh}, in the similar way, we find $%
\delta \left( n\right) =\pi $, and the quantization condition reads
\begin{equation}
\int_{0}^{x_{R}}\kappa dx=n\pi ,n=1,2,3...,
\end{equation}%
where $x_{R}=-1/E$. Eq.(14) gives the exact energy eigenvalues $%
E_{n}=-1/\left( 2n^{2}\right) $. For the three-dimensional Coulomb potential
$V_{eff}\left( r\right) =-1/r+l\left( l+1\right) /2r^{2}$, we find $\delta
\left( n,l\right) =\left[ \sqrt{l\left( l+1\right) }-l\right] \pi $, and the
quantization condition reads
\begin{equation}
\int_{r_{L}}^{r_{R}}\kappa dr+\left[ \sqrt{l\left( l+1\right) }-l\right] \pi
=\left( n+1\right) \pi ,n=0,1,2...,
\end{equation}%
which determines the exact energy eigenvalues $E=-1/\left( 2N^{2}\right) $
with $N=l+n+1$.

Eqs.(13)-(16) are novel and immediate expressions under no other auxiliary
modifications, whose exactness can make one convinced that the
scattering-led phase shift $\delta \left( n,l\right) $ does exist
objectively, and it plays a crucial role in determining the energy
eigenvalues. Because of it, when the WKB approximation is applied to the
three-dimensional potentials, Langer-correction is needed. Interestingly, if
the number $l\left( l+1\right) $ is instead of $\left( l+1/2\right) ^{2}$ in
Eq.(14) and Eq.(16), they are naturally changed to Langer-correction
formalism $\int_{x_{L}}^{x_{R}}\sqrt{2m\left[ E-V_{eff}\left( r\right) %
\right] }dr=\left( n+1/2\right) \pi $ with $V_{eff}\left( r\right) =V\left(
r\right) +\left( l+1/2\right) ^{2}/2r^{2}$ . Moreover, when the Coulomb-like
potentials is tackled, the problem of singularity is naturally avoided. With
regard to the unsolvable potentials, generally $\delta \left( n,l\right) $
is not an analytical expression of $n$ and $l$.
\begin{table}[tbp]
\begin{tabular}{cccc}
\hline\hline
$n$ & $E^{exact}$ & $E^{present}$ & $E^{SWKB}$ \\ \hline
0 & 14.312470 & 14.312470 & 14.312670 \\
1 & 14.468822 & 14.468822 & 27.174940 \\
2 & 42.227424 & 42.227424 & 39.499957 \\
3 & 43.899629 & 43.899629 & 47.940588 \\
4 & 68.213256 & 68.213256 & 62.315504 \\
5 & 74.975448 & 74.975448 & 76.652144 \\
6 & 94.567829 & 94.567829 & 92.325304 \\
7 & 108.464359 & 108.464359 & 108.985251 \\
8 & 124.328691 & 124.328691 & 126.433812 \\ \hline\hline
\end{tabular}%
\caption{Comparisons of exact, present, and SWKB energy eigenvalues for the
double oscillator $V\left(x\right)=10\left(|x|-3\right)^{2}$, setting $%
m=\hbar =1 $.}
\end{table}
Next the energy eigenvalues for the unsolvable radial Woods-Saxon
potential\cite{11} and the unsolvable double oscillator
potential\cite{12} with Eq.(12) are tabulated in Table I and II,
respectively, and the complete
agreement with exact results can be seen, implying Eq.(12) is exact. $%
E^{exact}$ is obtained through numerical technique. Dealing with the double
oscillator potential, as the energy of the particle is relatively smaller,
four turning points can occur, assuming $x_{1}<x_{2}<x_{3}<x_{4}$. Because
the particle will be motioning in the range $[x_{1},x_{4}]$, the
quantization condition $\int_{x_{1}}^{x_{4}}\left( \kappa -\kappa ^{^{\prime
}}P/P^{^{\prime }}\right) dx=\left( n+1\right) \pi $ gives the energy
eigenvalues. For other complex potentials with more then four turning
points, in the similar way, the contour is taken from the smallest turning
point to the biggest turning point. In these cases, the tunneling effect
takes place.

In summery, based on the transfer matrix rather than the wave function
expansion in powers of $\hbar$, we have derived the exact quantization
condition for arbitrary nonrelativistic quantum systems and have shown it
unifies and transcends the historical approaches. Naturally, we have
obtained the corresponding real quantum action-angle variable, contrasted
with the classical Hamilton-Jacobi theory, which discloses the
quantum-classical correspondence.

This work is supported by National Natural Science Foundation of P.R.China
under grant No.60237010 Municipal Scientific and Technological Development
Project of Shanghai under grant No.012261021, 01161084.


\begin{thebibliography}{99}
\bibitem{langer} R. E. Langer, Phys. Rev. 51, 669(1937).
\bibitem{miller} S. C. Miller and R. H. Good, Phys. Rev. 91,174(1953).
\bibitem{firedrich} H. Friedrich and J. Trost, Phys. Rev. Lett. 76, 4869(1996).
\bibitem{comtet} A. Comtet, A. D. Bandrauk, and D. K. Campbell, Phys. Lett. B 150, 159(1985).
\bibitem{barclay} D. T. Barclay, A. Khare and U. Sukhatme, Phys. Lett. A 183, 263(1993).
\bibitem{cao} Z. Q. Cao, Y. Jiang, Q. S. Shen, X. M. Dou and Y. L. Chen,
J. Opt. Soc. Am. A 16, 2209(1999).
\bibitem{sch} L. S. Schulman, \emph{Technigues and Applications of Path Integration}
(Wiley, New York, 1987).
\bibitem{33} P. Gaspard, D. Alonso, and I. Burghardt,
Adv. Chem. Phys. 90, 105(1995).
\bibitem{bohm} D. Bohm, Phys. Rev. 85, 166(1952); 85, 180(1952).
\bibitem{de} O. F. de Alcantara Bonfim, Phys. Rev. E 58, 2693(1998).

\bibitem{jh} J. Hainz and H. Grabert, Phys. Rev. A 60, 1698(1999).

\bibitem{11} E. Koch, Phys. Rev. Lett. 76, 2678(1996).
\bibitem{12} B. Chakrabarti and T. K. Das, Phys. Rev. A 60, 104(1999).

\end{thebibliography}
\end{document}